\documentstyle[aps,prl,multicol]{revtex}
\input{epsf}
\begin{document}
\draft
\preprint{\today}
\title{New state of the matter between the Fermi glass and 
the Wigner crystal \\ in two dimensions}
\author{Giuliano Benenti, Xavier Waintal and Jean-Louis Pichard}
\address{CEA, Service de Physique de l'Etat Condens\'e,
           Centre d'Etudes de Saclay, 91191 Gif-sur-Yvette, France}
\maketitle
\begin{abstract} 
 Spinless fermions with Coulomb interaction in a square disordered 
lattice form a new state of the matter which is nor a Fermi glass, 
neither a Wigner crystal for intermediate Coulomb interaction. 
From a numerical study of small clusters, we find that this new state 
occurs between the two critical carrier densities where 
insulator-metal-insulator transitions of a hole gas have been 
observed in GaAs.
\end{abstract}

\begin{multicols}{2}
\narrowtext

%
%

 An important parameter for a system of charged particles is the Coulomb 
energy to Fermi energy ratio $r_s$. In a disordered two-dimensional 
system, the ground state is obvious in two limits.  For large $r_s$, 
the charges form a pinned Wigner crystal, the Coulomb repulsion 
being dominant over the kinetic energy and the disorder. For small 
$r_s$, the interaction becomes negligible and the ground state is a Fermi 
glass with localized one electron states, in agreement with the scaling 
theory of Anderson localization. There is no theory for intermediate $r_s$, 
while many transport measurements following the pioneering works of 
Kravchenko et al~\cite{kravchenko} and made with electron and hole gases 
give evidence of an intermediate metallic phase in two dimensions, 
observed~\cite{hamilton} for instance when $6 < r_s < 9$ for a hole gas 
in GaAs heteostructures. Our study of spinless fermions with Coulomb 
repulsion in small disordered $2d$ clusters confirms that there is a new 
ground state for those values of $r_s$. In a given cluster, as we turn 
on the interaction, the Fermi ground state can be followed from $r_s=0$ 
up to a first level crossing at $r_s^{F}$. A second crossing occurs at a 
larger threshold $r_s^{W}$ after which the ground state can be followed 
to the limit $r_s \rightarrow \infty$. There is then an intermediate state 
between $r_s^{F}$ and $r_s^{W}$. In small clusters, the location of the 
crossings depends on the considered random potentials, but a study over the 
statistical ensemble of the currents supported by the ground state 
gives us two well defined values for $r_s^{F}$ and $r_s^{W}$: Mapping the 
system on a torus threaded by an 
Aharonov-Bohm flux, we denote respectively $I_{l}$ and $I_{t}$ the total 
longitudinal (direction enclosing the flux) and transverse parts of the 
driven current. One 
finds for their typical values $|I_{t}/I_{l}| \approx \exp -(r_s/r_s^{F})$ 
and $I_{l} \approx \exp-(r_s/r_s^{W})$ with $r_s^{F} < r_s^{W}$. For the 
Fermi glass, the flux gives rise to a glass of local currents and the sign 
of $I_{l}$ can be diamagnetic or paramagnetic, depending on the random 
potentials. For the intermediate state $(r_s^{F} < r_s < r_s^{W})$, the 
transverse current is suppressed while a plastic flow of longitudinal 
currents persists up to $r_s^{W}$, where charge crystallization occurs. 
The sign of $I_{l}$ can be paramagnetic or diamagnetic depending on the 
filling factor (as for the Wigner crystal), but does not depend on the 
random potentials (in contrast to the Fermi glass). This suggests that a 
theorem~\cite{legett} giving the sign of the current in $1d$ could be extended 
in $2d$ when $r_s > r_s^{F}$.  For a disorder yielding Anderson 
localization inside the clusters, one finds $r_s^{F}\approx 5.8$ and 
$r_s^{W}\approx 12.4$ in agreement with the values given by transport 
measurements which we shortly review.

%
%
 In exceptionally clean GaAs/AlGaAs heterostructures, an insulator-metal 
transition (IMT) of a hole gas results~\cite{yoon} from an increase  
of the hole density induced by a gate. This occurs at $r_s \approx 35$, 
in close agreement to $r_s^{W}\approx 37$, where charge crystalization 
takes place according to Monte Carlo calculations~\cite{tanatar}, 
and makes highly plausible that the observed IMT comes from the quantum 
melting of a pinned Wigner crystal. The values of $r_s$ where an IMT 
has been previously seen in various systems  (Si-Mosfet, Si-Ge, GaAS) 
are given in Ref.~\cite{yoon}, corresponding to different degrees of 
disorder (measured by the elastic scattering time $\tau$). Those $r_s$ 
drop quickly from $35$ to a constant value $r_s \approx 8-10$ when $\tau$ 
becomes smaller. This is again 
compatible with $r_s^{W} \approx 7.5$ given by Monte Carlo 
calculations~\cite{chui} for a solid-fluid transition in presence of 
disorder. If the observed IMT are due to interactions, it might be 
expected that this metallic phase will cease to exist as the carrier 
density is further increased. This is indeed the case~\cite{hamilton} 
for a hole gas in GaAs heterostructures at $r_s \approx 6$ where an 
insulating state appears, characteristic of a Fermi glass with weak 
electron-electron interactions. A similar observation has been also 
reported~\cite{pudalov1} for electrons in Si-Mosfet at a lower 
$r_s \approx 2$. The second transition towards Fermi glass  
can only be easily seen if the disorder is sufficient for driving 
localization effects inside the phase breaking length, 
a condition which weakens the metallic behavior of the intermediate phase. 

%
%

   The purpose of this work is to take advantage of exact diagonalization 
techniques possible only for small clusters at low filling factors. One 
can question the ability of this method to give results valid for large 
systems. In the clean limit, Pikus and Efros~\cite{tanatar} have 
obtained $r_s^{W}\approx 35$ diagonalizing $6 \times 6$ 
clusters with $6$ particles, close to $r_s^{W}\approx 37$ obtained 
by Tanatar and Ceperley for the thermodynamic limit. This is a first reason 
to study how one goes from the Fermi glass towards the Wigner crystal in 
small clusters. A second justification will be given a posteriori by the 
agreement between our numerical results and the experiments. 
%
%
 We consider a two--dimensional model of Coulomb interacting 
spinless fermions in a random potential. It is defined on a 
square lattice with $L^2$ sites occupied by $N$ electrons. 
The Hamiltonians reads 
\begin{eqnarray} 
\label{hamiltonian} 
H=-t\sum_{<i,j>} c^{\dagger}_i c_j +  
\sum_i v_i n_i +\frac{U}{2} \sum_{i\neq j} \frac{1}{r_{ij}} (n_i n_j), 
\end{eqnarray} 
where $c^{\dagger}_i$ ($c_i$) creates (destroys) an electron in 
the site $i=(i_x,i_y)$, $\sum_{<i,j>}$ denotes a sum restrained to nearest 
neighbors, $t$ is the strength of the hopping terms characterizing 
the particle kinetic energy ($t=1$ in the following) and $r_{ij}$ 
is the inter-particle distance for a $2d$ torus (minimum image 
convention). The random potential $v_i$ of the site $i$ with occupation 
number $n_i=c^{\dagger}_i c_i$ is taken from a box distribution of width $W$. 
The strength $U$ of the true physical long-ranged electron-electron 
interaction, suitable for insulating phases where screening breaks 
down, gives a Coulomb energy to Fermi energy ratio 
$r_s=U/(2t\sqrt{\pi \nu})$ for a filling factor $\nu=N/L^2$. 
For $U=0$ the Hamiltonian (\ref{hamiltonian}) reduces 
to the Anderson model of localization, while for $t=0$ it reduces to the 
Coulomb glass model, which describes classical point charges in a random 
potential. Exact diagonalization techniques for large sparse matrices 
(Lanczos method) are used to study a statistical ensemble 
of small clusters. The boundary conditions are always taken periodic in 
the transverse $y$-direction, and are chosen periodic, antiperiodic, 
or such that the system becomes a torus enclosing an Aharonov-Bohm flux 
$\phi$ in the longitunal $x$-direction. 

\begin{figure}
\centerline{
\epsfxsize=8cm 
\epsfysize=11cm 
\epsffile{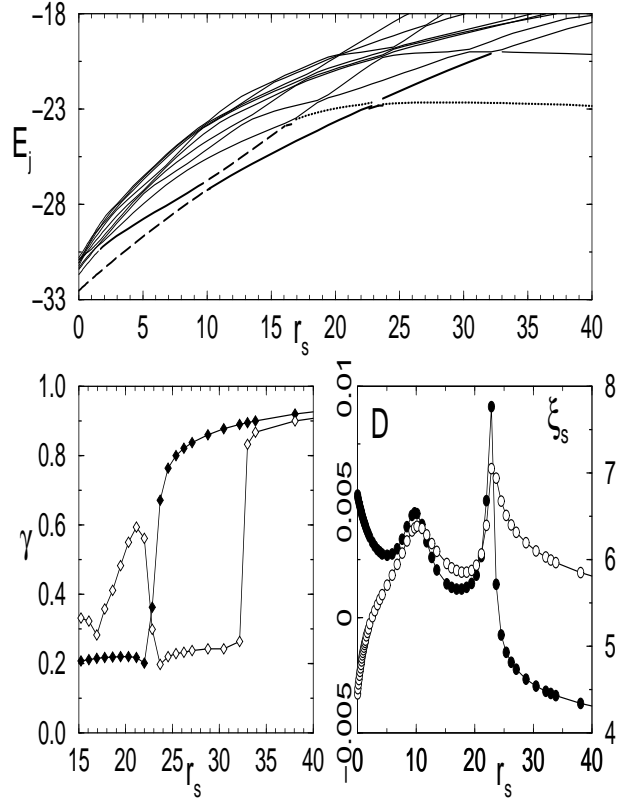}
}
\caption{Top: Low energy spectrum of a small single cluster 
($N=4$, $L=6$, $W=15$) (a $1.9 r_s$ term has been substracted); 
Bottom left: jumps of $\gamma$ at the second crossing where the 
ground state (filled diamonds) and the first excited state (empty 
diamonds) are interchanged. Bottom right: Ground state sensitivity 
$D$ (left scale, empty circles) and number of occupied sites $\xi_s$ 
(rigth scale, filled circles).}
\label{fig1} 
\end{figure}

%
  Fig.~\ref{fig1} exhibits behaviors characteristic of individual small 
clusters ($L=6$, $N=4$), where disorder ($W=15$) yields a strong 
Anderson localization, as a function of $r_s$ ($\nu=1/9$ and $0<U<50$). 
Looking at the low energy part of the spectrum, one can see that, as 
we gradually turn on the interaction, classification of the levels 
remains invariant up to first avoided crossings, visible for the 
ground state at $r_s^{F}$, where a Landau theory of the Fermi glass 
is certainly no longer possible. Looking at the electronic density 
$\rho_i = \langle \Psi_0| n_i| \Psi_0 \rangle$ of the ground state 
$|\Psi_0\rangle$, we have noticed that it is mainly maximum in the 
minima of the site potentials for the Fermi glass, with restrictions 
coming from kinetic terms and Pauli principle. After the second avoided 
crossing at $r_s^{W}$, $\rho_i$ is negliglible except for four sites 
forming a lattice of charges as close as possible to the Wigner crystal 
triangular network in the imposed square lattice. 
The degeneracy of the crystal is removed by the disorder, the array being 
pinned in $4$ sites of favorable energies. The $\rho_i$ of the intermediate 
ground state give the impression of a mixed phase with 
crytalline domains co-existing with glassy domains where fermions 
occupy the sites of lowest energies. 
  
For the same cluster, we have calculated the density-density correlation 
function $C(r)=N^{-1} \sum_i \rho_i \rho_{i-r}$ and the parameter 
$\gamma$ used by Pikus and Efros~\cite{tanatar} for characterizing the 
melting of the crystal. The parameter 
$\gamma=\max_{\,r} C(r) - \min_{\,r} C(r)$ 
calculated for the ground state and the first excited state around $r_s^{W}$ 
allows us to identify the second crossing with the melting of the crystal, 
since $\gamma=1$ for a crystal and $0$ for a liquid. Moreover, 
one can see that the Wigner crystal becomes unstable in the intermediate 
phase, while the ground state is related to the first excitation of the 
Wigner crystal above $r_s^{W}$ (Fig.~\ref{fig1} bottom left). The general 
picture is reminiscent to those found~\cite{sjwp} in one dimensional models. 
As we increase $r_s$, there are level crossings associated to charge 
reorganizations of the ground state. As in ref.~\cite{sjwp}, those charge 
reorganizations are accompanied by noticeable delocalization effects of the 
ground state (Fig.~\ref{fig1} bottom right). This is shown by a sharp  
enhancement of the ground state sensitivity $D=E(0)-E(\pi)$ measuring the 
change of the ground state energy when the boundary conditions are twisted 
in the $x$-direction, i.e. more fundamentally the ability of the ground state 
to a support a persistent current when the system forms a torus enclosing 
a flux. The second quantity demonstrating the delocalization effect at the 
crossings is the participation ratio $\xi_s =N^2( \sum_i \rho_i^2)^{-1}$ of 
the ground state, i.e. the number of sites that it occupies. Fig.\ref{fig1} 
is representative of the ensemble, with the restriction that the location 
of the crossings fluctuates from one sample to another, as 
observed~\cite{sjwp} in $1d$, as well as the sign (paramagnetic 
or diamagnetic) of $D$ below $r_s^{F}$, in contrast to $1d$. 

%
%

 Keeping the boundary condition periodic in the transverse $y$-direction  
and such that the system encloses a flux $\phi$ (in dimensionless units) 
in the longitudinal $x$-direction ($\phi=\pi$ coresponds to anti-periodic 
condition), one drives a persistent current of total longitudinal and 
transverse components given by:
$$
I_{l}=-\frac{\partial E(\phi)}{\partial \phi}=\frac{\sum_i I_i^x}{L_y} 
\ {\rm and} \ I_{t}=\frac{\sum_i I_i^y}{L_x}
$$ 
respectively. The local current flowing at the site $i$ in the 
$x$-direction is defined by 
$$
I_i^x=2 {\rm Im} (\langle \Psi_0 | c^{\dagger}_{i_{x+1},i_y} 
c^{}_{i_x,i_y} | \Psi_0 \rangle)
$$ 
and by a corresponding expression for $I_i^y$. The response is 
paramagnetic if $I_{l}>0$ and diamagnetic if $I_{l} < 0$. Legett 
theorem~\cite{legett}, based on a simple variational possiblity for 
the ground state wave function 
$\Psi_0$, states that the sign of the response is given by the parity of 
$N$. $(-1)^N (E(0)-E(\pi))$ is always positive in $1d$ for all 
disorder and interaction strength. The proof is based on the 
nature of ``non symmetry dictated nodal surfaces '', which is 
trivial in $1d$, but which has a quite complicated topology in higher $d$. 
Moreover, in $2d$ and small $r_s$, the sign of $I_{l}$ depends on 
the site potentials.
%
%
 
\begin{figure}
\centerline{
\epsfxsize=8cm 
\epsfysize=4cm 
\epsffile{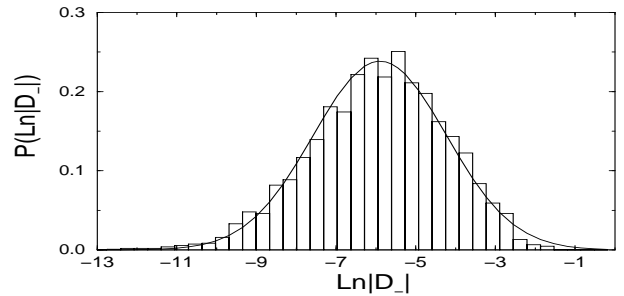}
}
\caption{Distribution of the ground state sensitivity $D_{-}$ of the 
diamagnetic samples of an ensemble of $10^4$ clusters at $r_s=1.7$ 
fitted by a log-normal.}
\label{fig2} 
\end{figure}

\begin{figure}
\centerline{
\epsfxsize=8.5cm 
\epsfysize=9.5cm 
\epsffile{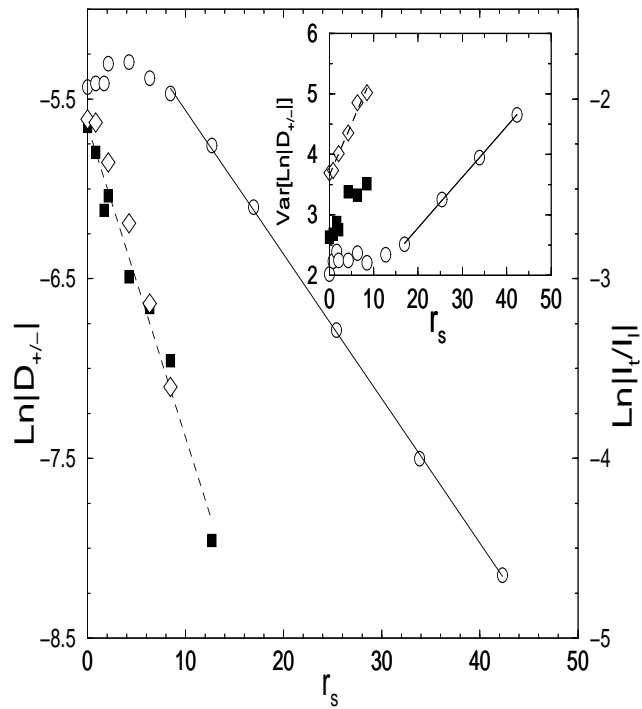}
}
\caption
{Log-averages of $D_+ \propto \exp-({r_s/12.4})$ (paramagnetic, 
empty circle) and $|D_-| \propto \exp-({r_s/5.8})$ (diamagnetic, 
filled square) given by the left scale, and of the ratio $|I_{t}/I_{l}|$ 
(empty diamond, right scale). Insert: variances of $\log |D_{-}|$, 
of $\log |I_{t}/I_{l}| \propto r_s/5.9$ and of $\log |D_{+}| \propto  
r_s/11.9$.
}
\label{fig3} 
\end{figure}

 We have separately studied the paramagnetic and diamagnetic samples 
of an ensemble of $10^3$ clusters with again $L=6$, $N=4$ and $W=15$.
We obtain lognormal distributions for all values of $r_s$ for a 
large disorder ($W=15$), 
as illustrated by Fig.\ref{fig2}. The dependence on $r_s$ of 
the averages and variances of the logarithms of the paramagnetic 
$D_{+}=E(0)-E(\pi) >0 $ and diamagnetic $D_{-}=E(0)-E(\pi) < 0$ 
responses are given in Fig.\ref{fig3}.  The log-averages exponentially 
decay as $D_{-} \propto \exp-(r_s/r_s^{F})$ and $D_{+} \propto 
\exp-(r_s/r_s^{W})$ with $r_s^{F} \approx 5.8$ and $r_s^{W}\approx 12.4$. 
The decay of the typical values are accompanied by an increase of their 
fluctuations. The $r_s$-dependence of the log-variances 
are shown in the insert, showing that the variances and the averages 
are proportional. This reminds us a similar relation~\cite{pichard} 
for the log-conductance in a $2d$ Fermi glass at $r_s=0$. The variances of 
$\log |I_{t}/I_{l}|$ and $\log D_{+}$ increase as $r_s/r_s^{F}$, and as 
$r_s/r_s^{W}$ above $r_s^{W}$ respectively. Both the average and the 
variance of $\log|D_{+}|$ remain nearly constant in the intermediate phase. 

 A transition from glassy towards plastic flow as $r_s$ increases has been 
found~\cite{avishai} by Berkovits and Avishai. It is likely that the 
existence of diamagnetic samples comes from disordered arrays of local 
loops of current which should give rise to a transverse current $I_{t}$. 
Studying the distribution of $I_{t}$ and $I_{l}$ at $\phi=\pi/2$, we find 
that there is indeed a non zero transverse current $I_{t}$ when $r_s$ is 
small. Its sign is random over the statistical ensemble, and the 
ratio $\log |I_{t}/I_{l}|$ is normally distributed. The log-average plotted 
in Fig.\ref{fig3} confirms that $|I_{t}/I_{l}| \propto \exp-(r_s/r_s^F)$ 
exactly as $D_{-}$. 

 The fact that $I_{l}$ becomes paramagnetic independently on the 
microscopic disorder when $r_s>r_s^{F}$ does not mean that  
Coulomb repulsions always yield this response. For instance, 
$4\times6$ clusters with $N=6$ particles become always diamagnetic 
when $r_s$ increases. One just concludes that there is a rule for  
$r_s>r_s^{F}$ giving the sign of the response in $2d$, as the one 
found by Legett in $1d$, which does not depend on the random potential.
\begin{figure}
\centerline{
\epsfxsize=8cm 
\epsfysize=8cm 
\epsffile{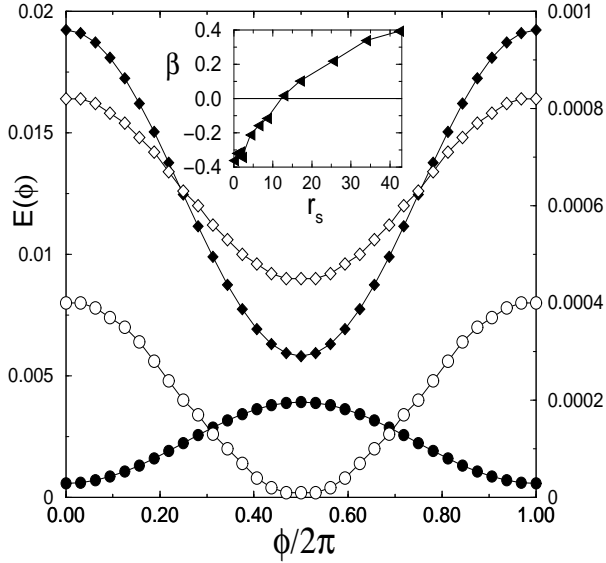} 
}
\caption{Flux dependence of ground state and first excited state 
energies for $r_s=0$ (filled symbols, left scale) and $r_s=42$ 
(empty symbols, right scale) for a single cluster. A suitable 
energy translation has been done to put the curves in the same 
figure. Insert: ensemble average of $\beta=D_0D_1/|D_0D_1|$.  
}
\label{fig4} 
\end{figure} 
  It is also interesting to study the currents supported by the first 
excited state. In a Fermi glass, if the ground state current $I_{l}(E_0)$ 
has a certain sign, the current $I_{l}(E_1)$ carried by the first 
excitation has very often the opposite sign. We have noticed that 
the opposite behavior becomes more likely when $r_s$ increases: 
the currents carried by the ground state and the first excited state 
flow in the same direction. An illustration is given in Fig.~\ref{fig4}, 
where the flux dependence of the two first levels at $r_s=0$ and $r_s=42$ 
of the same cluster is shown. The transition from anti-correlated 
towards correlated flux dependence can be seen on the average of 
$D_0D_1/|D_0D_1|$ which precisely changes its sign at $r_s^{W}$, 
as shown in the insert. Fig.\ref{fig4} also confirms that the 
approximation $-{\partial E_j}/{\partial \phi} \approx D_j=E_j(0)-E_j(\pi)$ 
makes sense. 
%
%

  In summary, we have shown that a simple model of spinless fermions 
with Coulomb repulsion in a random potential can account for the 
two critical densities of holes where insulator-metal-insulator 
transitions occur. This gives an important confirmation that the 
intermediate phase observed in two dimensions results from 
Coulomb long-range repulsions, and makes unlikely scenarios based on  
spin-orbit scattering~\cite{pudalov2} or on some special single electron 
interface properties~\cite{altshuler}. In contrast to Finkelstein's 
approach~\cite{finkel}, we underline that the spins are neglected in 
our model, and we cannot confirm that the intermediate ground 
state is a metal for a disorder energy to kinetic energy ratio as large 
as $W/t=15$. We have mostly seen noticeable delocalization effects near the 
crossing points. We cannot exclude that weaker disorders or spin effects are 
necessary for having a metal. We also note that the possibility of a 
superconducting state has been discussed~\cite{phillipps}, and not 
entirely ruled out in the clean limit~\cite{tsiper}.
A study of the temperature dependence of the ground state sensitivity 
and of the system size dependence of the gap excitations at a fixed $r_s$ 
will be useful for clarifying this issue. From these small cluster studies, 
we conclude that there is a new state of the matter, clearly separated 
from the Fermi glass and from the Wigner crystal, identified by a plastic 
flow of currents and a magnetic response with a sign independent on the 
microscopic disorder. The obtained critical $r_s$ factors are surprisingly 
close to those given by Ref.~\cite{hamilton}. However, the critical $r_s$ can 
depend on $W$, as already known for $r_s^{W}$, and might have 
small finite size corrections (a study of $D_{-}$ in $8\times 8$ clusters 
with $N=4$ gives $r_s^{F}\approx 6.2$ instead of $5.8$).

 A partial support by the TMR network ``Phase coherent 
dynamics of hybrid nanostructures'' of the EU is gratefully acknowledged.

\end{multicols}

\begin{references}

\bibitem{kravchenko} 
S.V.~Kravchenko et al. {\it Phys.\ Rev.} B {\bf 50}, 8039 (1994);
ibid {\bf 51}, 7038 (1995).  

\bibitem{hamilton}
A.R.~Hamilton, M.Y.~Simmons, M.~Pepper, E.H.~Linfield, P.D.~Rose 
and D.A.~Ritchie, preprint cond-mat/9808108.

\bibitem{legett}
A.J.~Legett, {\it Granular Nanoelectronics}, D.K.~Ferry ed., p 297, 
Plenum Press, N.Y. (1991).

\bibitem{yoon}
J.~Yoon, C.C.~Li, D.~Shahar, D.C.~Tsui and M.~Shayegan, 
preprint cond-mat/9807235.

\bibitem{tanatar}
B.~Tanatar and D.M.~Ceperley, {\it Phys.\ Rev.} B {\bf 39}, 5005 (1989); 
F.G.~Pikus and A.L.~Efros, {\it Solid State Commun.} {\bf 92}, 485 (1994).

\bibitem{chui}
S.T.~Chui and B.~Tanatar, {\it Phys.\ Rev.\ Lett.} {\bf 74}, 458 (1995).

\bibitem{pudalov1}
V.M.~Pudalov, G.~Brunthaler, A.~Prinz and G.~Bauer, preprint cond-mat/9812183.

\bibitem{sjwp} 
P.~Schmitteckert, R.A.~Jalabert, D. Weinmann and J.-L.~Pichard, 
{\it Phys.\ Rev.\ Lett.} {\bf 81}, 2308 (1998). 

\bibitem{pichard}
S.~Feng and J.L.~Pichard, {\it Phys.\ Rev.\ Lett.} {\bf 67}, 753 (1991).

\bibitem{avishai}
R.~Berkovits and Y.~Avishai, {\it Phys.\ Rev.} B {\bf 57}, R15076 (1998). 

\bibitem{pudalov2}
V.M.~Pudalov, {\it JETP Lett.} {\bf 66}, 175 (1997).

\bibitem{altshuler}
B.L.~Altshuler and D.L.~Maslov, {\it Phys.\ Rev.\ Lett.} in press.

\bibitem{finkel}
A.M.~Finkelstein, {\it Sov.\ Sci.\ Rev} A {\it Phys} {\bf 14}, 3 (1990). 

\bibitem{phillipps}
P.~Phillipps et al., {\it letters to nature} {\bf 395}, 254 (1998).

\bibitem{tsiper}
E.V.~Tsiper and A.L.~Efros, {\it Phys.\ Rev.} B {\bf 57}, 6949 (1998).

\end{references}
\end{document}